\documentclass[prb,twocolumn,preprintnumbers,amsmath,amssymb,floatfix]{revtex4-1}
\usepackage[dvipsnames]{xcolor}

\usepackage{graphicx}
\usepackage{dcolumn}
\usepackage{bm,textcomp}
\usepackage{booktabs,threeparttable}
\usepackage{subfigure}
\usepackage{epstopdf}
\usepackage{color}
\makeatletter 
\def\@hangfrom@section#1#2#3{\@hangfrom{#1#2#3}}
\makeatother
\makeatletter 
\def\@biblabel#1{[#1]}
\makeatother

\newcommand{\cm}[1]{} 
\DeclareGraphicsExtensions{.eps,.mps,.pdf,.jpg,.png}
\definecolor{bg}{rgb}{0.75,.75,.67}

\begin{document}

\title{Formation and dilatation of shear bands in a Cu-Zr metallic glass: A free volume perspective}
\author{Chunguang Tang$^{1*}$, Hailong Peng$^{2}$, Yu Chen$^{1}$, Michael Ferry$^1$}

\affiliation{$^1$ School of Materials Science and Engineering, University of New South Wales, NSW 2052, Australia; $^2$ Institut f\"{u}r Materialphysik im Weltraum, Deutsches Zentrum f\"{u}r Luft- und Raumfahrt (DLR),
51170 K\"{o}ln, Germany}

\begin{abstract}

We study the tensile deformation behaviour of metallic glass Cu$_{50}$Zr$_{50}$ as a function of quenching rate using molecular dynamics simulations. The atomic scale shearing is found to be independent on atomic free volume, and the macroscopic correlation between the yield strength and the density (or average free volume) is a coincidence that the samples with large free volume also have a low density of shear-resistant local five-fold symmetry. In the relatively slowly quenched ($\leq 10^{10}$ K/s) samples, shear bands have a dilatation about 0.5\%, which compares well with recent experiment results. In contrast, although more active local shearing occurs in the rapidly quenched samples, shear bands are not observed. This is because the strain energy disperses into local atomic shearing at the macroscopically elastic stage and, hence, is not sufficient for shear band activation, resulting in homogeneous deformation and appreciable plasticity.
\end{abstract}
\maketitle
\section{Introduction}

Bulk metallic glasses (BMGs) are among the most promising metallic materials, partly because of their ultrahigh strength due to their glassy structure generated by rapid quenching from the liquid phase. However, most monolithic BMGs usually suffer catastrophic failure without noticeable macroscopic plastic deformation. This is due to the localization of plastic strain in a few shear bands that propagate in an autocatalytic manner, leaving regions outside the bands to be only elastically deformed. Suppression of major shear band formation and propagation is crucial for  improving the mechanical properties of BMGs. Factors determining the formation and evolution of shear bands have attracted intense research interest in the metallic glass community. The recent progress on shear band research has been extensively reviewed.\cite{greer_shear_2013,wang_source_2015}  

Shear bands are often conceived to nucleate from some structural ``defects'' such as free volume\cite{spaepen_microscopic_1977,spaepen_homogeneous_2006}, which is the free space around atoms facilitating irreversible atomic rearrangement. This local irreversible process often involves a group of atoms or the so-called shear transformation zones (STZs)\cite{argon_plastic-deformation_1979} that cooperatively accommodate plastic strain in the surrounding elastic matrix. Thus, free volume is conventionally considered to be the main factor triggering local plastic events in metallic glasses. However, this model was challenged recently.\cite{wang_source_2015} The free volume concept, initially created for hard-sphere systems, assumes that diffusion and flow are microscopically realized by hopping of individual atoms into adjacent vacancy sites (i.e., free volume).\cite{spaepen_microscopic_1977} In reality, however, atoms can squeeze into non-vacancy space due to the nearly harmonic interatomic potential and the associated diffusion process can occur via a collective motion of a group of atoms.\cite{wang_collective_1999} Therefore, nowadays the free volume concept commonly refers to the location of low density which may not contain a vacancy. Generally, free volume may affect the mobility of atoms. For example, in CuZr alloys a correlation between density and diffusivity has been found.\cite{zhang_effects_2014}

As the nucleation of shear bands in BMGs occurs on an extremely small scale, their direct experimental observation is difficult, although the size of STZs has been estimated indirectly at the elastic-plastic transition by nanoindentation.\cite{choi_estimation_2012} This renders computer simulations a favourable alternative approach for studying the atomistic mechanisms of shear banding in BMGs. A number of simulations have estimated the size of STZs, which  is a complex function of the external loading conditions and material properties, to range from tens to a few hundred atoms.\cite{srolovitz_atomistic_1983,zink_plastic_2006, mayr_activation_2006}  The activation energies for STZs have also been estimated using different simulation methods \cite{greer_shear_2013} and range from 0.35 eV for a STZ of size $\sim$140 atoms\cite{mayr_activation_2006} to 0.3 eV for 10$\sim$20 atoms, \cite{delogu_molecular_2008} which translates to 0.003 to 0.03 eV/atom.

Not only are the formation dynamics of shear bands complicated, their thermodynamic properties are also far from being clear. For example, there remains some debate about the atomic volume change in a shear band region as shear bands form and propagate, although this is central to any free volume-based deformation theory. A shear deformation simulation of Cu$_{64}$Zr$_{36}$ has reported a shear band dilatation of 2-3\% at the peak stress and about 6\% in the mature shear band.\cite{li_free_2007} In a tension simulation\cite{cao_structural_2009} of this system the volume increase of Cu atoms in a mature shear band, which is clearly distinguishable from the volume fluctuation in the matrix, is around 10\% in our estimation, whereas another tension study reported a volume increase of only 0.3\% for Cu and 2.3\% for Zr.\cite{feng_atomic_2015} For other systems, a small volume increase up to 1\% was reported for Mg$_{85}$Cu$_{15}$, which was comparable to the intrinsic atomic volume fluctuations in the matrix.\cite{bailey_atomistic_2006} Experimentally, early transmission electron microscopy studies \cite{donovan_structure_1981} on Fe$_{40}$Ni$_{40}$B$_{20}$ argued a large volume increase of the order of 10\% in the shear bands, but a recent examination of a single shear band produced in Zr$_{69.5}$Cu$_{12}$Ni$_{11}$Al$_{7.5}$ reported the most extreme dilatation of only 1.14\%.\cite{pan_softening_2011} An $in~situ$ acoustic emission measurement indicates a range of dilatation with the most probable value below 2\%.\cite{klaumuenzer_probing_2011}

Herein, we perform molecular dynamics simulations of the tensile deformation of metallic glass Cu$_{50}$Zr$_{50}$ systematically as a function of quenching rate in an attempt to elucidate any relationship between quenching rate, local structures, and shear band formation. We also study the dilatation in shear bands via monitoring the change of atomic volumes.

\begin{figure}[htb]
\centering
\includegraphics[width=3.2in]{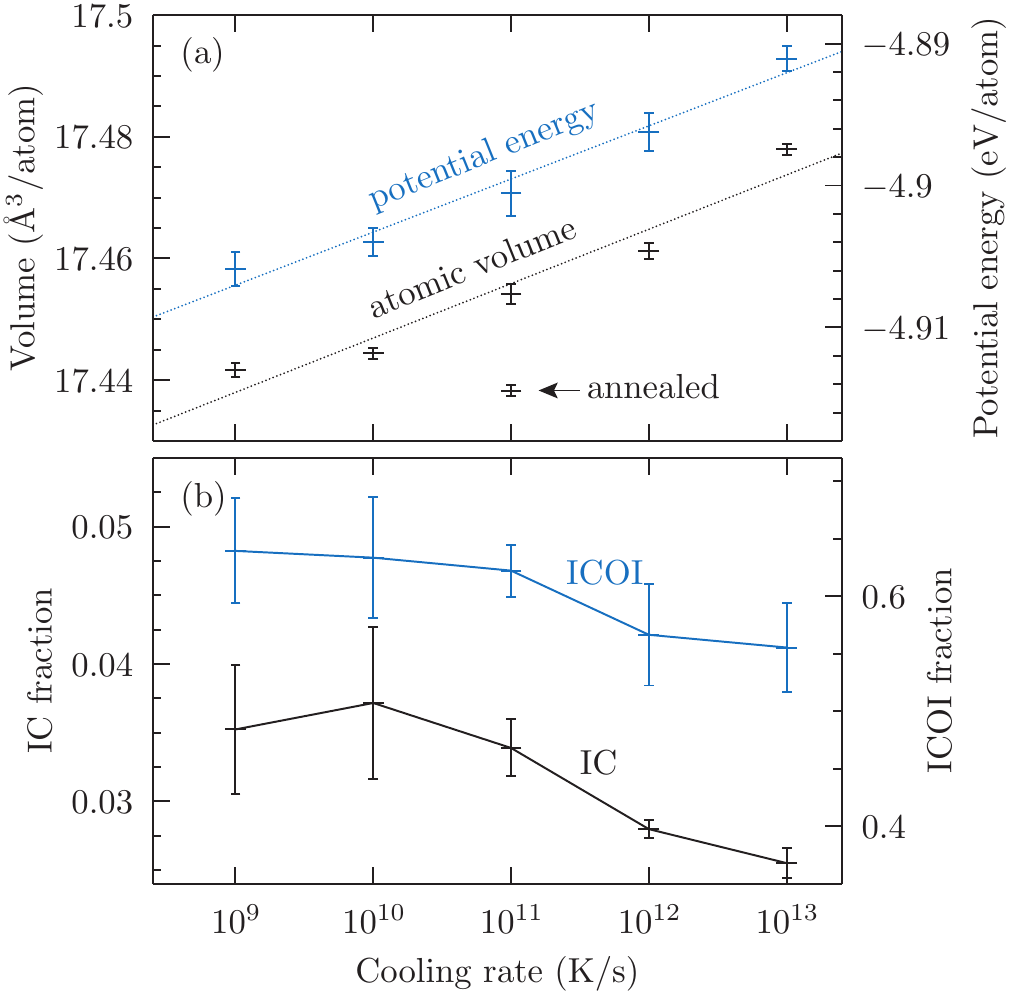}
\caption{Effect of cooling rate on the average atomic volume and the potential energy of CuZr at 100 K. The volume of annealed sample 10$^{11}$ K/s is also shown.}
\label{fig:rate-vol}
\end{figure}

\begin{figure}[htb]
\centering
\includegraphics[width=3.2in]{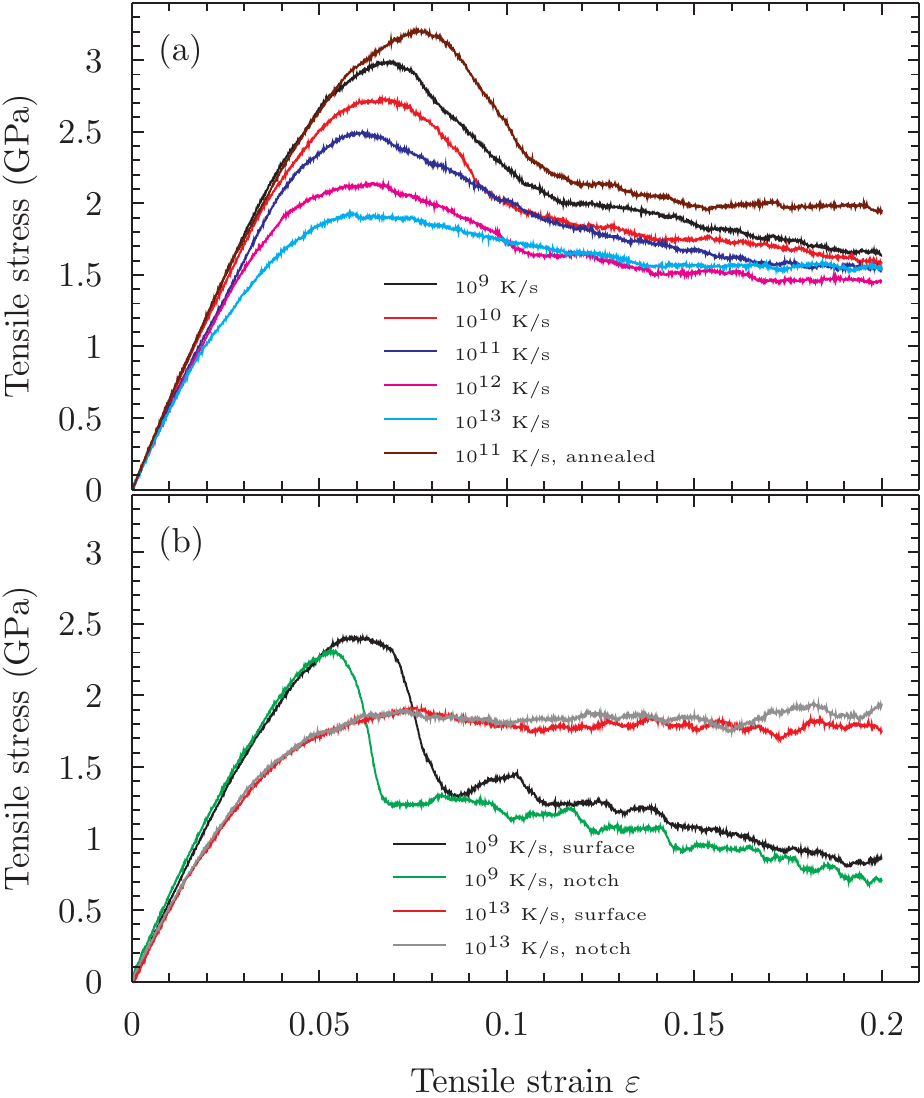}
\caption{(a) Stress-strain curves for CuZr quenched at different rates followed by straining under periodic boundary conditions. The ultimate strength increases with decreasing quenching rate, and is highest for the annealed sample. (b) Stress-strain curves for samples with free surfaces or with a notch on one of the free surfaces. Note that the tests were performed with strain constrained.}
\label{fig:strstr}
\end{figure}

\begin{figure*}[htb]
\centering
\includegraphics[width=6in]{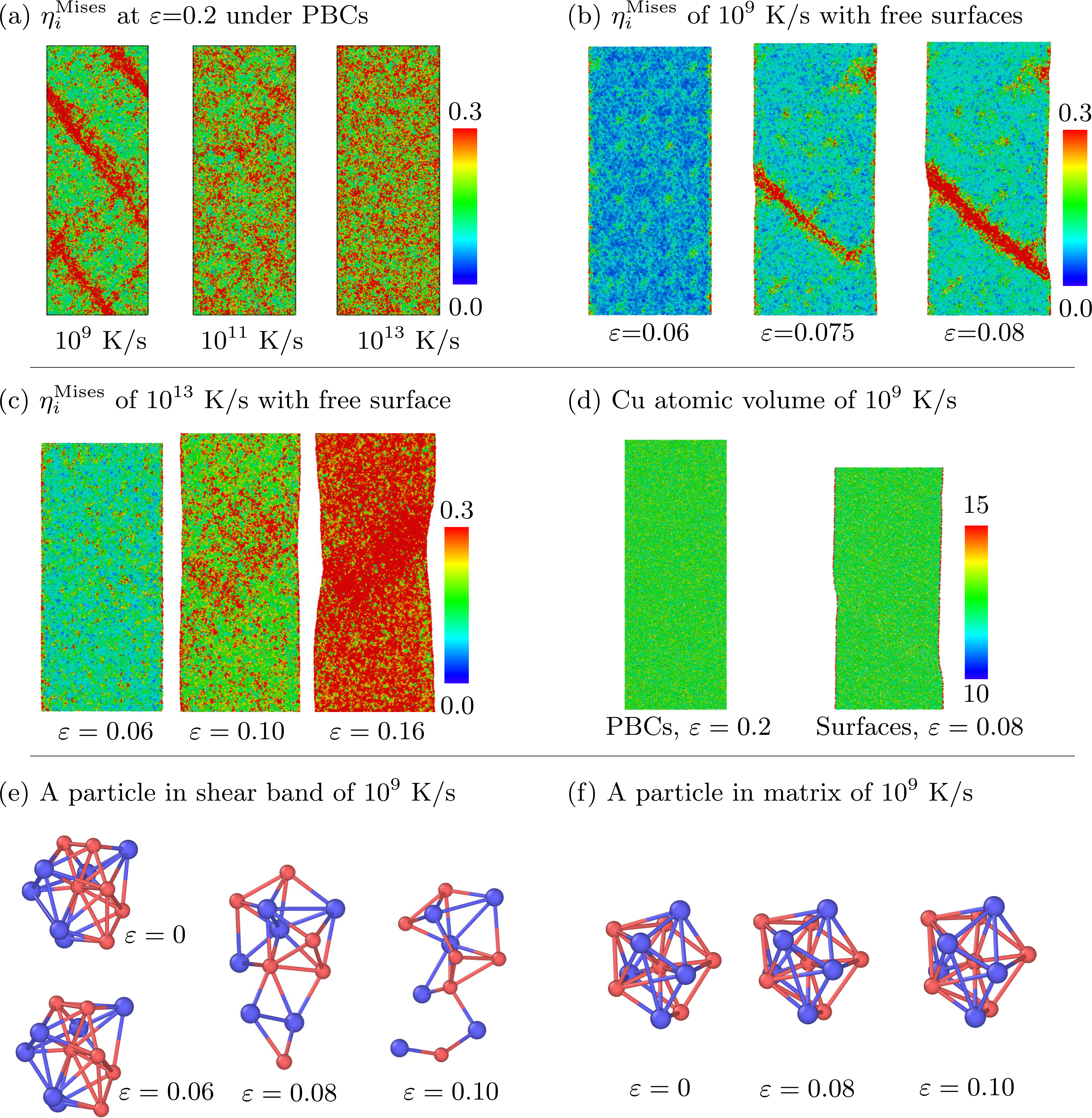}
\caption{(a-c) Distribution of atomic strain in samples with atoms colored according to their shear strain $\eta^{\rm{Mises}}_{i}$. Color map is for atomic strain and the slabs shown in (b) and (c) are sliced through the slab centers. (d) Color map for Voronoi volume of Cu atoms in sample 10$^9$ K/s. The sample with free surfaces was sliced to remove the surface atoms of which the Voronoi volume cannot be defined meaningfully. The color code is for volume between 10 and 15 $ $\AA$^3$. (e-f) Typical cluster shape change in sample 10$^9$ K/s with surfaces as a function of tensile strain.}
\label{fig:shearband}
\end{figure*}

\begin{figure}[!htb]
\centering
\includegraphics[width=3.2in]{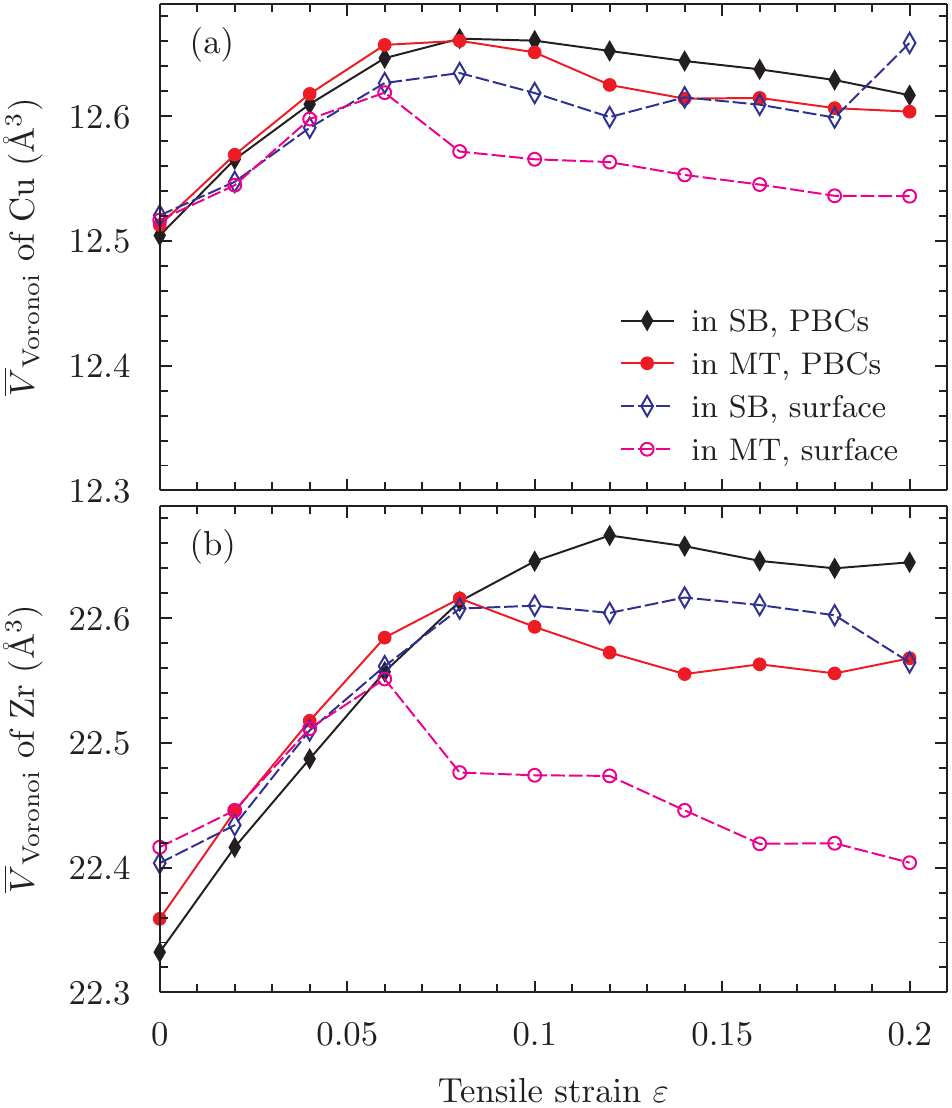}
\caption{Average Voronoi volume of Cu and Zr in shear bands (SB) and matrix (MT) of sample 10$^9$ K/s. As shear bands form at large tensile strains, the average volumes of both Cu and Zr become larger in the shear bands than in the matrix. The standard deviations of the data are about 0.6 for Cu and 0.8 for Zr, respectively, for both samples under periodic boundary conditions (PBCs) and with surfaces.}
\label{fig:vorovolume}
\end{figure}

\begin{figure*}[htb]
\centering
\includegraphics[width=6in]{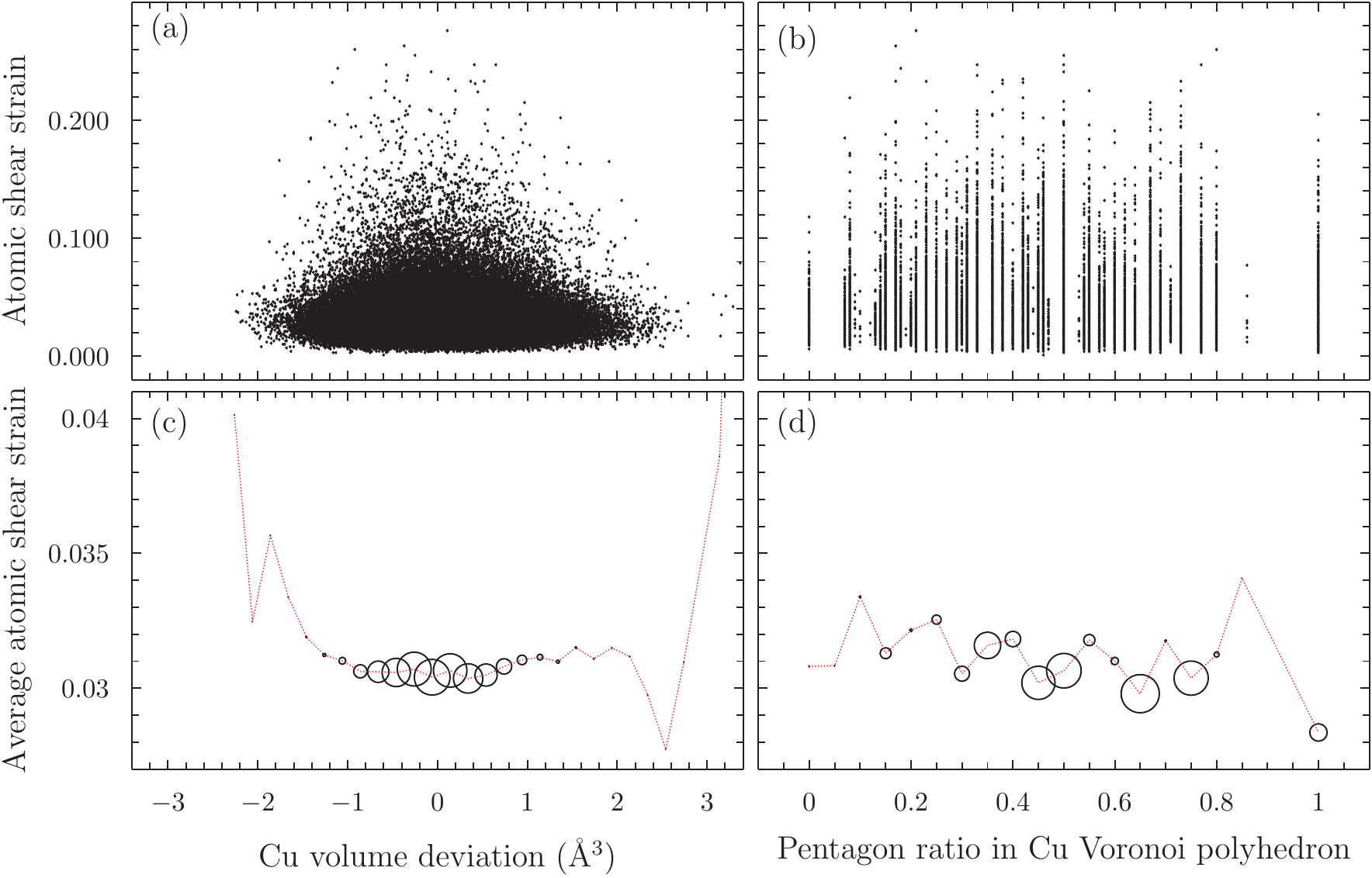}
\caption{Relationship among atomic shear strain $\eta^{\rm{Mises}}_{i}$, Voronoi volume, and pentagon fraction of Voronoi polyhedra for Cu atoms in sample 10$^9$ K/s at tensile strain $\varepsilon$=0.08 under periodic boundary conditions. A reference configuration 10 picoseconds before $\varepsilon$=0.08 was used for $\eta^{\rm{Mises}}_{i}$ calculation, and volume and pentagon fraction are for the reference configuration. The results for Zr atoms are qualitatively similar. (a-b) Distribution of $\eta^{\rm{Mises}}_{i}$ and atomic volume or pentagon fraction. One data point for each atom. (c-d) Average $\eta^{\rm{Mises}}_{i}$ for atoms within a Voronoi volume slot (width 0.2 $ $\AA$^3$) or a pentagon fraction slot (width 0.05). The diameters of the circles represent their probability and the dashed lines are for eye guide only.}
\label{fig:vol-5fold-strain}
\end{figure*}

\section{Method}
The simulations were carried out as follows: A bulk cubic simulation cell containing 6750 atoms was relaxed at 2000 K for 2 ns to make the system well liquefied, followed by cooling to 100 K at five quenching rates, i.e., 10$^{9}$, 10$^{10}$, 10$^{11}$, 10$^{12}$, and 10$^{13}$ K/s (referred throughout as sample 9, 10, 11, 12 and 13). The final structural configurations were used to construct the tensile test samples by repeating the building blocks $1\times5\times10$ in $x$, $y$, and $z$ directions, respectively. The samples were relaxed at 100 K for 2 ns before subjecting them to strain-constrained tension at a strain rate of 10$^8$/s.  All simulations were carried out under NPT (constant particle number, pressure, and temperature) ensemble with periodic boundary conditions (PBCs) in $x$, $y$, and $z$ directions. For quenching rates 10$^{9}$ and 10$^{13}$ K/s, tensile samples with free surfaces (parallel to $z$ axis) and surface notches were also constructed and tested under NPT ensemble. All simulations were carried out using code LAMMPS,\cite{lammps} with the atomic interactions modelled by the embedded atom method (EAM) potential proposed by  Mendelev $et~al$.,\cite{Mendelev2009} and post-production visualization and Voronoi analysis were performed using codes OVITO\cite{stukowski_visualization_2010} and VORO$++$.\cite{rycroft_voro_2009} During Voronoi analysis, we assumed the atomic radii of Cu and Zr to be 1.26 and 1.58 $ $\AA$ $,\cite{ward_structural_2013} respectively.

The atomic scale deformation is often characterized by the local shear invariant $\eta^{\rm{Mises}}_{i}$ proposed by Shimizu $et~al$\cite{Shimizu2007} or the minimum non-affine squared displacement $D^2_{min}$ by Falk and Langer.\cite{falk_dynamics_1998} In both cases, the strain computation for some atom $i$ involves counting the squared displacements of all the neighbouring atoms, $D^2$, relative to atom $i$. A suitable affine transformation matrix $\mathbf{J}$ is introduced to minimize $D^2$ and effectively remove the elastic portion of the displacement, leaving $D^2$ an indicator of only the plastic deformation. The local shear invariant $\eta^{\rm{Mises}}_{i}$ is then derived from the Lagrangian strain tensor which is a function of ${\mathbf{J}}$. Both $D^2_{min}$ and $\eta^{\rm{Mises}}_{i}$ are good measures of local inelastic deformation and here we characterized the deformation using $\eta^{\rm{Mises}}_{i}$.

\section{Results and discussion}

We begin with checking the effect of quenching rate on the density and stability of samples. As the quenching rate decreases on a logarithmic scale, the density increases roughly linearly, from 7.35 g/cm$^3$ for sample 13 to 7.37 g/cm$^3$ for sample 9, or in terms of average atomic volume, from 17.48 to 17.44 $ $\AA$^3 $, as shown in Fig. \ref{fig:rate-vol}. A similar trend was found for the energetic stability. Since atoms cannot be compressed or stretched, the average atomic volume is also an indicator of the average free volume in the system. As density increases, the yield strength of the samples also increases (Fig. \ref{fig:strstr}(a)), and this demonstrates a correlation between the shearing instability and the average free volume. To confirm this correlation we reduced the free volume in sample 11 by annealing at 650 K, which is slightly below the glass transition temperature,\cite{zhang_effects_2014} for 400 ns and found that its strength increased to a level corresponding to its new density.

The stress-strain curves result from a collective effect from all the atoms, but for observing the nucleation and propagation of shear bands we are more interested in some local events. To this end we plot in Fig. \ref{fig:shearband}(a) the distribution of atomic shear strain $\eta^{\rm{Mises}}_{i}$ for some selected quenching rates at tensile strain $\varepsilon=0.2$. Shear bands were observed in samples 9 and 10 (not shown), but not in the samples with higher quenching rates. Our results are supported by a previous simulation report\cite{cao_structural_2009} that shear bands were not observed in metallic glass samples with ultra-fast cooling. It is interesting to note that the annealed sample 11 (not shown) also exhibits clear shear bands.

The above results are for samples with PBCs in $x$, $y$, and $z$ directions, but experimental samples have free surfaces and probably stress concentrators such as voids and surface notches, which would facilitate the plastic deformation. For completeness, we performed tensile tests on samples 9 and 13 in the presence of free surfaces and surface notches. As shown in Fig. \ref{fig:strstr}(b), for sample 9 with surfaces and/or a notch the peak stress reduces significantly and a sudden stress drop occurs following the peak stress, whereas for sample 13 the influence of surface and notch on the stress is negligible. Consistent with these stress-strain data, for sample 9 with free surfaces, a shear band nucleates around $\varepsilon=0.07$ and propagates through the sample at $\varepsilon=0.08$, as can be seen from Fig. \ref{fig:shearband}(b), significantly earlier than in the PBCs case, and the shear band forms even earlier in the case of surface notch (not shown). For sample 13, however, the introduction of free surfaces or surfaces plus a notch does not induce shear bands and only necking occurs at large strain (Fig. \ref{fig:shearband}(c)).

Fig. \ref{fig:shearband}(d) shows the Voronoi volume of Cu in sample 9 with PBCs at tensile strain $\varepsilon=0.2$ and with free surfaces at $\varepsilon=0.08$. At these strains,  mature shear bands have formed but the volume color maps of the samples are featureless. Similar featureless maps for Zr atoms were observed. We noted that the Voronoi volume of Cu and Zr in this system fluctuates approximately within 11-14 and 20-25 $ $\AA$^3$, respectively. Hence, it is possible that the atoms in the shear bands experience some small dilatation which is indistinguishable in the volume map due to the volume fluctuation. In order to confirm this, we chose some atoms in the shear band zones and the matrix, respectively, and measured the average volume of these atoms as a function of the tensile strain. 

As shown in Fig. \ref{fig:vorovolume}, for the sample under PBCs, before $\varepsilon=0.1$ the average atomic volumes in the two zones are similar and increase with $\varepsilon$ and, after that, the volume in shear band zones becomes larger than in the matrix, which is more obvious for Zr species. For the sample with free surfaces, similar trends exist and the volume difference is already obvious at $\varepsilon=0.08$. The average volumes in the shear band zones reach a relatively stable stage after the shear band formation, and the further applied external strain results in sample sliding along the shear bands (in the free-surface case). It is interesting to note that the atomic volumes in the matrix zones decrease to some extent, especially for the free-surface case, implying relaxation in the matrix upon the shear band formation. 

It should be noted that the atomic volume is dynamically modulated during local shearing operation. For those atoms with very large positive or negative free volume, they are energetically unfavorable and may decrease or increase their volume through deformation, resulting in free volume annihilation or creation. The dilatation of shear bands is thus a result of the dynamic process of both free volume creation and annihilation. Overall, we estimate the dilatation of the shear bands to be 0.5\% for the free-surface case. This number translates to a radial increase of about 0.16\% in the atomic space, or not more than 0.01 $ $\AA$ $ increase in the first peak position of the radial distribution function, which is hardly detectable, as supported by a recent study\cite{sepulveda-macias_onset_2016} on Cu$_{50}$Zr$_{50}$. Our result is comparable with the simulations \cite{bailey_atomistic_2006, feng_atomic_2015} and the recent measurement,\cite{pan_softening_2011} but is significantly smaller than those reported in other simulations.\cite{li_free_2007, cao_structural_2009} We note that in one of the latter cases,\cite{li_free_2007} the reported dilatation of about 6\% is observed at an applied shear strain as large as 40\%, and at the maximum stress point, corresponding to a shear strain of about 20\% where the shear band is already distinct from the matrix, the dilatation is only about 2-3\%. The relatively larger dilatation observed in this work is probably due to the sample under shear deformation experiencing large shearing in the shear band without failure. 
 
While we did not measure the softening resulting from this small dilatation, the recent work\cite{pan_softening_2011} on a single shear band reported a significant softening of 36\% from the dilatation of 1.14\%, which implies that even a very small dilatation is enough for shear bands to propagate in an autocatalytic manner.

While the stress-strain curves suggest that the shearing instability increases with the average free volume increase, in a single sample we have not observed such a correlation on the atomic scale, that is, ${\eta}^{\rm{Mises}}_{i}$ for a specific atom $i$ exhibits no simple linear dependence on its (free) volume. For example, for sample 9 we plotted the volume and ${\eta}^{\rm{Mises}}_{i}$ for each atom in Fig. \ref{fig:vol-5fold-strain}(a) and no clear relationship can be found. On the other hand, it has been proposed that the proneness of an atom to shearing is related to its local five-fold symmetry, or the fraction of pentagon facets in the Voronoi polyhedron.\cite{peng_structural_2011} Again, we have not found a clear relationship between the two parameters for a specific atom, as shown in Fig. \ref{fig:vol-5fold-strain}(b). However, the average ${\eta}^{\rm{Mises}}_{i}$ appears to strongly correlate with the five-fold symmetry, becoming smaller with higher symmetry (Fig. \ref{fig:vol-5fold-strain}(d)). While in general the average ${\eta}^{\rm{Mises}}_{i}$ does not correlate with the atomic volume, as also being reported for mono-atomic metallic glasses,\cite{srolovitz_atomistic_1983} we find that those atoms with volumes far deviated from the average one tend to have larger ${\eta}^{\rm{Mises}}_{i}$, as shown in Fig. \ref{fig:vol-5fold-strain}(c). This can be understood since these atoms, either with large positive or negative free volume, are in higher potential energy states and thus less mechanically stable compared with other atoms. 

In view of the lack of a microscopic monotonic correlation between the atomic shear strain and the free volume, the macroscopic monotonic correlation between the strength and the average free volume (or density) remains to be explained. One possibility is that the atomic volumes of the rapidly quenched samples are much larger than those of the slowly quenched ones such that most atoms of the former are in high energy states and, hence, prone to shearing. However, this is not the case as the distributions of atomic volumes in samples 9 and 13 are found to be qualitatively very similar.

\begin{figure}[!htb]
\centering
\includegraphics[width=3.1in]{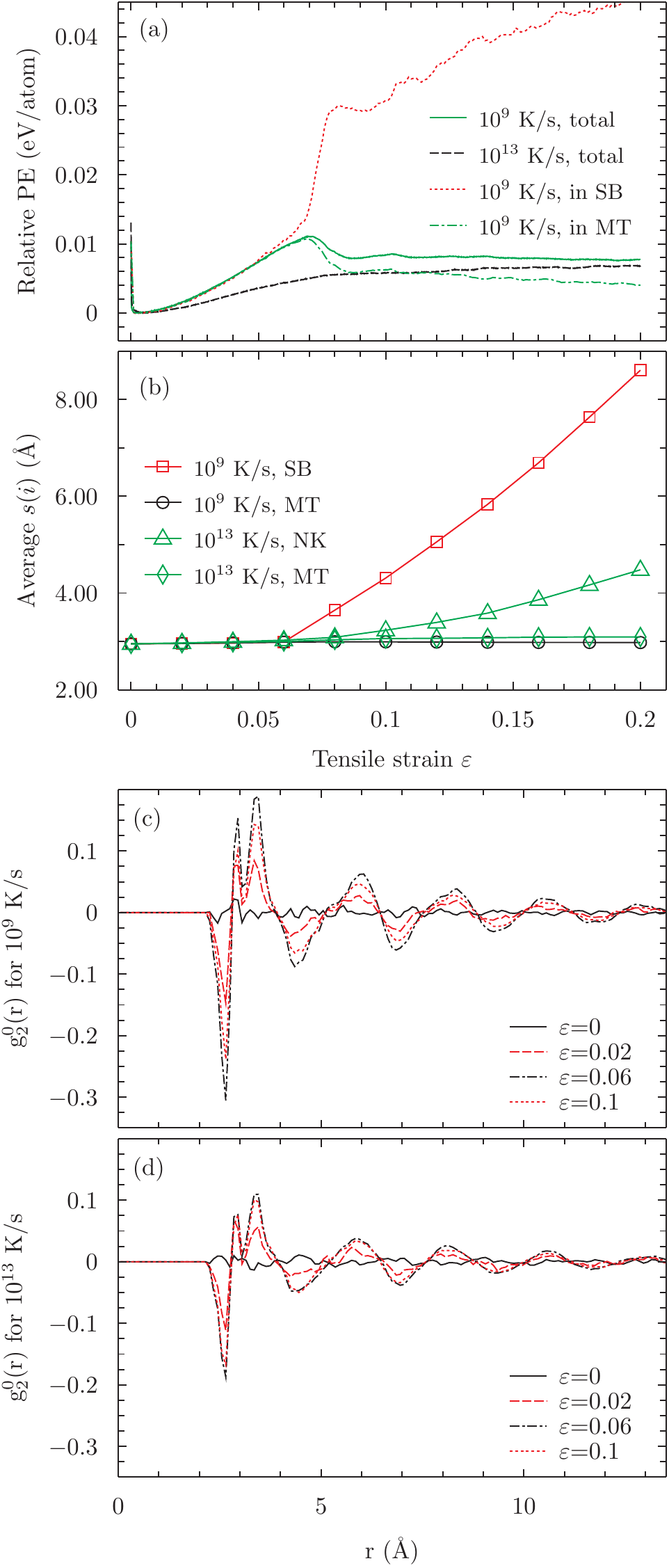}
\caption{(a) Atomic potential energy in samples 10$^9$ and 10$^{13}$ K/s with free surfaces. `SB' for shear band and `MT' for matrix. (b) Average distorted ``size'' of clusters in samples with free surfaces. The ``size'' is defined as the average distance between the central atom and its neighbouring atoms.  `NK' for necking area. Only bulk atoms are selected for consideration. (c-d) Component $g^0_2$($r$) of the pair distribution function for samples 10$^9$ and 10$^{13}$ K/s at various tensile strains under periodic boundary conditions. }
\label{fig:hetero}
\end{figure}

Our Voronoi analysis indicates that, as the quenching rate increases, not only does the average free volume increase, the fraction of pentagons in the Voronoi polyhedra also decreases, because in rapid quenching, there is less time for the atoms to relax into the energetically more stable icosahedron-like polyhedra, which possess a high fraction of pentagons. Interestingly, even at the same quenching rate, the correlation between the average free volume and pentagon fraction was also found in Cu-Zr alloys with a range of compositions.\cite{wakeda_relationship_2007} Since the rapidly quenched samples contain more polyhedra with low pentagon symmetry, macroscopically they tend to yield at lower stresses.

The formation of shear bands depends on the percolation of STZs, at least in the case of PBCs.\cite{greer_shear_2013} Since the shearing of the rapidly quenched atoms are more easily activated, one may expect easy formation of more STZs and, hence, shear bands in the rapidly quenched samples. However, the activation energies of STZs for shear banding are higher than those of atomic shearing. Moreover, as a result of atomic shearing spreading over the sample, the applied energy disperses into the sheared atoms and the threshold activation energy required for STZs is difficult to reach, similar to the case of shear band multiplication that retards the propagation of the main band. In contrast, in the slowly quenched samples, high elastic energy can be stored to trigger STZ formation. 

To test the above argument, we computed the potential energies in samples 9 and 13 with surfaces, as shown in Fig. \ref{fig:hetero}(a). In sample 9 the potential energy averaged over the whole sample reaches a maximum when shear band propagation begins ($\varepsilon\approx0.07$), then drops till the shear band passes through the sample ($\varepsilon\approx0.08$), and then stays relatively stable when the sample slides along the shear band. The potential energy stored in the sample by tension is greater than 0.01 eV/atom before shear band formation. In contrast, in sample 13 the potential energy increases by about 0.005 eV/atom at $\varepsilon\approx0.07$, and up to a stable $\sim$0.007 eV/atom at large strains where sample necking begins. The less increase in potential energy in sample 13 indicates that atoms are able to relax more in response to the applied strain, which is consistent with the lower strength of sample 13. 

A close examination of sample 9 shows that the potential energy in the shear band region starts to deviate from that in the matrix at $\varepsilon\approx0.05$, well before the shear band begins to propagate. We attribute this deviation to the activation of some STZs and estimate the activation energy to be $\sim$0.008 eV/atom, which falls in the range 0.003$\sim$0.03 eV/atom mentioned earlier.\cite{mayr_activation_2006,delogu_molecular_2008} During shear band propagation (around 0.07$<\varepsilon<$0.08) the potential energy in the shear band region increases sharply, probably due to the dilatation, but the energy in the matrix decreases, implying the strain/stress in the matrix is somewhat released. For $\varepsilon>0.08$ the energy rise in the shear band region is due to the increasing number of surface atoms as the sample slides along the shear band.

The above analysis proves that, energetically, a shear band is more difficult to be activated compared with necking. Structurally, this implies the local atomic environment changes more significantly in the shear band zone than in the necking zone. To quantify this difference, we use a simple parameter $s$($i$) to characterize the distorted ``size'' of cluster $i$, or the average distance between atom $i$ and a group of atoms that are the neighbours of atom $i$ at zero tensile strain. As shown in Fig. \ref{fig:hetero}(b), the clusters in the shear band zone experience more severe deformation than those in the necking zone. In the matrix, the size of the clusters experience little change. Typical pictures of cluster distortion in both the shear band and the matrix are shown in Fig. \ref{fig:shearband}(e-f).

The lack of shear bands in the rapidly quenched samples indicates that the atomic shear strain is more homogeneously distributed in these samples. From a structural perspective, the bond distribution in these samples is also more isotropic in response to the applied strain in tension. To illustrate this, we projected the pair distribution function $g$($\mathbf{r}$) onto the spherical harmonic functions $Y^{m}_{l}$($x$) as $g$($\mathbf{r}$)$=\Sigma_{l,m}g^m_l$($r$)$Y^m_l$($\mathbf{r}/r$). In this formula the isotropic component $g^0_0$ is the commonly used radial distribution function $g$($r$), which we found to be insensitive to the quenching rate and strain. Specifically, we compared the component $g^0_2$ along the loading axis for samples 9 and 13 as this component is a good indicator of the stress-induced anisotropy.\cite{dmowski_elastic_2010, peng_stress-versus_2013} As shown in Fig. \ref{fig:hetero}(c-d), for both samples the pair distributions become anisotropic in response to the applied strain, but sample 13 remains relatively more isotropic. 

\section{Summary}
We have simulated the tensile deformation of metallic glass Cu$_{50}$Zr$_{50}$ prepared with cooling rates from 10$^9$ to 10$^{13}$ K/s under three deformation situations: with periodic boundary conditions, with free surfaces along the loading axis, and with notches on the free surfaces. Some observations emerged from this work are: (a) Microscopically, atomic shearing is independent of atomic free volume for the majority of atoms, although a small fraction of atoms with very large positive or negative free volume tend to exhibit large atomic shearing because they are energetically metastable. The commonly observed correlation between the yield strength and the density is a coincidence that the samples with large free volume due to rapid quenching also have a low density of shear-resistant local five-fold symmetry; (b) Shear bands are only observed in samples quenched at rates of 10$^{10}$ K/s and lower, and their formation is facilitated by free surfaces and surface notches. In the rapidly quenched samples, the strain energy disperses into the easy local atomic shearing at the early stage, leaving not enough strain energy for shear band activation and resulting in homogeneous deformation and appreciable plasticity. The deformation of the rapidly quenched samples is insensitive to defects such as surfaces and notches; (c) The dilatation in shear bands is found to be about 0.5\%, which is smaller than some previous simulation results\cite{li_free_2007,cao_structural_2009} but compares well with some more recent experiment and simulation results.\cite{pan_softening_2011,klaumuenzer_probing_2011,feng_atomic_2015} 

\section{Acknowledgements}
The authors acknowledge NCI National Facility for computational support of project codes eu7 and y88. Chunguang Tang would particularly like to thank the Australian Research Council for the DECRA Fellowship (grant no. DE150100738) for enabling this work to be carried out.



\end{document}